\documentclass[
superscriptaddress,
twocolumn,
showpacs,
bibnotes,
amsmath,
amssymb,
aps,
pra,
floatfix,
]{revtex4-2}

\usepackage{graphicx}
\usepackage{dcolumn}
\usepackage{bm}
\usepackage{cleveref}
\usepackage{silence}
\usepackage{color}
\usepackage{soul}
\usepackage{amsmath}

\begin{document}


\title{Assessing the Reliability of Anomalous Hall Conductivity Extraction in GdAlSi} 
\author{Anil Kumar}
    
    \affiliation{Department of Physics and Astronomy, Texas Tech University, Lubbock, Texas 79409, USA.}

\author{Debapratim Pal}
    
    \affiliation{Department of Physics and Astronomy, Texas Tech University, Lubbock, Texas 79409, USA.}

\author{Sudhan Koirala}
    
    \affiliation{Department of Physics, New Mexico State University, Las Cruces, New Mexico 88001, USA.}

\author{Muhammad Adnan}
    
    \affiliation{Department of Physics and Astronomy, Texas Tech University, Lubbock, Texas 79409, USA.}

\author{Youngsang (Eric) Ji}
    
    \affiliation{Department of Physics and Astronomy, Texas Tech University, Lubbock, Texas 79409, USA.}

\author{Prakash Regmi}
    
    \affiliation{Department of Physics and Astronomy, Texas Tech University, Lubbock, Texas 79409, USA.}
    
\author{Bailey S. Bouley}
    
    \affiliation{Department of Chemistry and Biochemistry, Texas Tech University, Lubbock, Texas 79409, USA.}
    
\author{Ludi Miao}
    
    \affiliation{Department of Physics, New Mexico State University, Las Cruces, New Mexico 88001, USA.}

\author{Yun Suk Eo}
    \email{yeo@ttu.edu}
      
    \affiliation{Department of Physics and Astronomy, Texas Tech University, Lubbock, Texas 79409, USA.}

\date{\today}

\begin{abstract}
  We examine the different methods of extracting the anomalous Hall conductivity using SrRuO$_3$ and GdAlSi. For SrRuO$_3$, where the ordinary Hall background is well defined and the magnetoconductance is small, subtracting the ordinary Hall contribution either before or after conversion from resistivity to conductivity yields nearly identical anomalous Hall conductivities. In GdAlSi, in contrast, the transverse response does not exhibit clear saturation, and both low- and high-field regions can appear approximately linear. We show that different interpretations of these linear regions as the ordinary Hall background, together with the treatment of the measured longitudinal resistance in the resistivity-to-conductivity conversion, can produce completely different estimates of the anomalous Hall conductivity. We also examine whether the nonlinear Hall response can instead be described within an ordinary Hall framework without invoking an anomalous contribution. Our analysis provides practical cautions for extracting excess Hall contributions in quantum materials. 

\end{abstract}

\maketitle

\section{Introduction}

The Hall effect is one of the most widely used charge transport probes for studying quantum materials. The simplest case is the ordinary Hall effect (OHE), which originates from the Lorentz force acting on charge carriers in an applied magnetic field\cite{hall1879new}, and it provides information about carrier type, density, and mobility \cite{sprinkart2024tutorial, hurd1972dynamics, yue2017towards, dunlap2019normal}. The measured transverse response may contain additional contributions that cannot be described solely by the OHE. Such excess Hall channels have been well identified in various magnetic systems, as well as topologically non-trivial systems. Excess Hall effects include the anomalous Hall effect (AHE), which can originate from intrinsic Berry curvature or extrinsic scattering mechanisms, and the topological Hall effect (THE), which is associated with the real space Berry phase of the noncoplanar magnetic textures\cite{nagaosa2010anomalous, li2022interplay}. To study these excess Hall channels, it is therefore critically essential isolating these excess Hall channels by subtracting the ordinary Hall background from the total measured transverse channel.

Revealing the excess Hall channel from the measured transverse resistance by subtracting the OHE is not always a trivial task. For special cases, a distinct magnetic-field dependence of the excess Hall contribution, such as showing hysteresis or saturation behavior consistent with the magnetization, in contrast to an ordinary Hall response that remains approximately linear in field, allows these contributions to be separated easily. However, for many other systems, this separation becomes particularly difficult when the transverse resistance does not saturate and shows no clear unique signatures other than a slight change in slope from the behavior expected for the ordinary Hall effect. Reviewing the literature, several different methods of OHE subtraction have been used to report the anomalous Hall resistivity, $\rho_{yx}^{AHE}$, or anomalous Hall conductivity, $\sigma_{yx}^{AHE}$, including subtraction of a linear Hall resistivity determined at low fields\cite{liu2018giant}, subtraction of a linear Hall resistivity determined at high fields\cite{yin2025magnetism}, and subtraction performed after converting the Hall resistivity to Hall conductivity\cite{manna2018colossal}. The use of these substantially different analysis procedures raises two important questions: 1) Do these different methods lead to different estimates of the AHE or THE? 2) If so, why? and which method should be used at each condition?

One notable example of such complexity, yet a particularly interesting system because of its rich physics, is GdAlSi. Different pictures have been proposed for its underlying magnetic ground state. First-principles studies by Laha $et$ $al$.\cite{Laha_GdAlSi}. and Nag \textit{et al}\cite{Nag_GdAlSi}. suggest a collinear antiferromagnetic ground state. In particular, Nag $et$ $al$.. further proposed GdAlSi as an altermagnet, a recently identified class of materials hosting zero net magnetization and momentum-dependent spin splitting at the same time\cite{AltermagnetSeminal}. Laha \textit{et al}., on the other hand, found that a spiral magnetic state closely competes in energy with the collinear ground state\cite{Laha_GdAlSi}. More recently, Nakano \textit{et al}. performed polarization-resolved resonant x-ray scattering measurements and identified an incommensurate cycloidal magnetic ground state, together with a field-induced noncoplanar multi-$Q$ magnetic texture\cite{nakano2026perfectly}. Other recent transport and thermodynamic measurements have further revealed closely spaced magnetic transitions near the Neel temperature ($T_N$), as well as multiple magnetic-field-induced transitions\cite{li2026field}.

The interpretation of experimental measurements of the Hall response in GdAlSi has similarly not reached a consensus. Laha $et$ $al$. reported an exceptionally large anomalous Hall conductivity of approximately $1310~\Omega^{-1}\mathrm{cm}^{-1}$ at 2 K and attributed to the momentum-space Berry curvature associated with Weyl nodes\cite{Laha_GdAlSi}, while also discussing the alternative possibility of a nontrivial magnetic texture below the N\'eel temperature. In contrast, Gong $et$ $al$.. analyzed the Hall response within a conventional single-carrier picture and reported a normal Hall effect without invoking an anomalous Hall contribution\cite{gong2024magnetic}. Given the contrasting conclusions of the magnetic ground state and the possible coexistence of several mechanisms contributing to the excess Hall response, one must be extra careful with interpreting the transverse resistance measurement of  GdAlSi. 

Motivated by these rich reports on GdAlSi, we examine the different commonly used extraction methods of finding the excess Hall channel from the measured transverse resistance data. We compare linear extrapolations from both low- and high- field regions in both transverse resistivity and transverse conductivity. As a comparison, we first present data and analysis of SrRuO$_3$, whose anomalous Hall response is well-established and signatures of AHE saturation can be clearly seen in the raw data. We then study GdAlSi, where the transverse resistance does not saturate and its origin is still debatable. Our results demonstrate that, when the ordinary Hall background is not independently constrained, and the origin of the longitudinal magnetoresistance is ambiguous, the magnitude, field dependence, and even the sign of the anomalous Hall conductivity can strongly depend on the details of the extraction process. In the following section, we first review the transport fundamentals.


\section{Hall-transport framework}

In this section, we establish the transport framework needed for understanding the rest of the following sections. We first note that throughout this work, we focus on the anomalous Hall contribution defined as the excess transverse response in addition to the ordinary Hall effect. Other excess Hall contributions, such as the topological Hall effect, can be analyzed in the same way introduced in this section.

To experimentally study the anomalous Hall contribution, the ordinary Hall response must first be correctly estimated from the measured transverse resistance signal. This separation, however, requires assumptions of the ordinary Hall response and how the ordinary and anomalous contributions combine. 

We begin with the Drude model of the ordinary Hall effect and longitudinal magnetotransport. Then, we discuss how the conductivity and resistivity tensors are related, how multiple conduction channels combine, and how these intrinsic transport coefficients are connected to experimentally measured resistances.

\subsection{Ordinary Hall Effect and Drude Magnetotransport\label{SubSec:Drude}}

The ordinary Hall effect originates from the Lorentz force acting on the charge carriers when a magnetic field is applied perpendicular to both the current and transverse voltage measurement direction (i.e., $\vec{B} = B \hat{z}$). We first consider the classical Drude model for a single channel and isotropic case (i.e., $\rho_{xx} = \rho_{yy}$ ). After determining $\rho_{xx}$ and $\rho_{yx}$ from the resistance measurements, the resistivity components can be expressed as

\begin{equation}
      \rho_{xx} = \frac{1}{n e \mu} = \frac{m}{n e^2 \tau}
      \label{eq:rho_xx}
\end{equation}
and 
\begin{equation}
     \rho_{yx} = -\rho_{xy} = s \frac{B}{ne}
\end{equation}
where, $n$ is the carrier density, $m$ is the effective mass, $\tau$ is the scattering time, $\mu$ ($= e \tau /m$) is the mobility, $s$ is the charge species (electrons being -1 or holes being +1). 

From the perspective of Ohm's law, $\tilde{\rho} \vec{J} =  \vec{E}$, $\rho_{xx}$ and $\rho_{yx}$ are matrix elements of $\tilde{\rho}$. In addition, the matrix inversion of $\tilde{\rho}$ is $\tilde{\sigma}$, and we can use the alternative form of Ohm's law, $\vec{J} =  \tilde{\sigma} \vec{E}$.
From this matrix inversion, we find

\begin{equation}
     \quad \sigma_{yx} = -\sigma_{xy}= - s\frac{ne\mu^2 B}{1 + \mu^2 B^2}
\end{equation}
and 

\begin{equation}
     \sigma_{xx} = \frac{n e \mu}{1 + \mu^2 B^2}, 
     \label{eq:sigma_xx}
\end{equation}

where $s$ is the sign of carriers. 

When more than one conduction channel is present, it is important to understand how their contributions combine. From an electronic band perspective, different Fermi pockets experience the same applied electric field, or equivalently the same electrochemical-potential gradient. On the other hand, each channel can carry a different current because its transport properties, such as the carrier density and mobility, may differ. For two independent conduction channels described by conductivity tensors ($\tilde{\sigma_1}$) and ($\tilde{\sigma_2}$), the total current density is therefore the sum of the current densities carried by the individual channels,

 \begin{equation}
     \vec{J} = \vec{J_{1}} + \vec{J_{2}} = \tilde{\sigma_{1}} \vec{E} + \tilde{\sigma_{2}} \vec{E}.
 \end{equation}

Alternatively, we can write $\vec{J} = \tilde{\sigma}_{total} \vec{E}$, where

 \begin{equation}
     \tilde{\sigma}_{total} = \tilde{\sigma}_{1} + \tilde{\sigma}_{2}.
 \end{equation}

In contrast, multiple scattering mechanisms within the same conduction channel add through their scattering \textit{rates} rather than their scattering \textit{times}. Note the scattering rate is inversely proportional to the scattering time. If two independent scattering mechanisms are characterized by scattering times, $\tau_{1}$ and $\tau_{2}$, the total scattering time is 

\begin{equation}
    \frac{1}{\tau_{total}} = \frac{1}{\tau_{1}} + \frac{1}{\tau_{2}}.
\end{equation}

We can generalize this to $N$ channels, and obtain

\begin{equation}
    \sigma_{xx}^{total} = \sum_{i}^{N}\frac{n_{i} e \mu_{i}}{1+\mu_{i}^2B^2}
\end{equation}

and 

\begin{equation}
    \sigma_{yx}^{total} = -\sum_{i}^{N}s_{i}\frac{ n_{i}e\mu_{i}^2B}{1+\mu_{i}^2B^2}.
\end{equation}

For the two-channel case ($N=2$), the Hall response can still appear approximately linear when both channels are in the low-field regime ($\mu_i B\ll 1$), and when the longitudinal conductivity typically dominates the Hall conductivity  ($|\sigma_{yx}|\ll\sigma_{xx}$). For the case where the channels have the same sign ($s= s_{1} = s_{2}$), expanding the Hall resistivity in powers of magnetic field gives

\begin{equation}
\begin{aligned}
    \rho_{yx}
    &= -\frac{\sigma_{yx}}
    {\sigma_{xx}^2 + \sigma_{yx}^2} \\
    &= R_{H}^{0}B
    \left[1 - (\mu_{\mathrm{eff}}B)^2 + \cdots \right]
    \approx R_{H}^{0}B .
\end{aligned}
\label{Eq:EffectiveTwoHall}
\end{equation}

where 

\begin{equation}
    R_{H}^{0} = s\frac{n_{1}\mu_{1}^2+n_{2}\mu_{2}^2}{e(n_{1}\mu_{1} + n_{2}\mu_{2})^2}
\end{equation}

and 

\begin{equation}
    \mu_{eff}^2= \frac{n_{1}n_{2}(n_{1}+n_{2}) \mu_{1}^2\mu_{2}^2 (\mu_{1}-\mu_{2})^2}{(n_{1}\mu_{1} + n_{2}\mu_{2})^2 (n_{1}\mu_{1}^2 + n_{2}\mu_{2}^2)}.
\end{equation}

For low-mobility channels, $\mu_{\mathrm{eff}}B\ll1$ can remain satisfied over a substantial magnetic-field range, causing the two-channel Hall response to appear nearly indistinguishable from the linear Hall response of a single-channel system.

Later we will use this Drude model to interpret the ordinary magnetotransport, including the ordinary Hall effect. In the following subsection, we turn our attention to what information is actually contained in a resistance measurement. In particular, we examine how the measured longitudinal and transverse resistances are related to the intrinsic resistivity and conductivity tensors, and how this relationship can depend on the sample geometry.

\subsection{Resistivity, Conductivity, and Measured Resistance \label{SubSec:ResistivityConductivity}}

\begin{figure}[h]
    \centering
    \includegraphics[width=1\columnwidth]{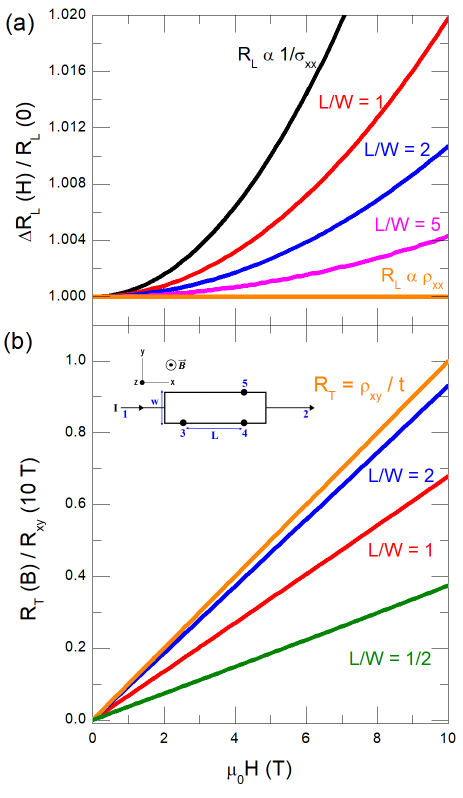} 
    \caption{Finite-element simulation of magnetotransport (single-channel Drude model). (a) Normalized longitudinal resistance ($R_L(B)/R_L(0)$) for different length-to-width ratios, together with the ideal cases ($R_L\propto 1/\sigma_{xx}$) and ($R_L\propto\rho_{xx}$). (b) Normalized transverse resistance (measured at the center of the sample) for different length-to-width ratios, together with the ideal case ($R_{T} = \rho_{yx}/t$) (Used $\rho_{xx}(0)=30~\mu\Omega \cdot \mathrm{cm}$ and $\mu=200~\mathrm{cm^2/V \cdot s}$ for the simulations). Inset: A typical longitudinal and transverse resistance configuration for experiments (electrodes enumerated). }
    \label{Fig:FEA_LongitudinalResistance}
    \end{figure}

In this subsection, we review how a resistance measurement is connected to the Drude model that we reviewed in the previous subsection. We will later show that when interpreting the anomalous Hall resistivity,  the correct interpretation of the ordinary magnetotransport (i.e., Drude model) becomes critical. A typical magnetotransport measurement looks like the inset in Figure~(\ref{Fig:FEA_LongitudinalResistance}) (b). A bar-shaped sample is prepared with several electrodes. Electrodes 1 and 2 are used to send electrical current. Electrodes 3, 4, and 5 are used to measure voltage. 

The longitudinal resistance ($R_{xx}$) can be obtained by sending the current from electrodes 1 to 2 ($I_{12}$) and measuring the voltage difference between electrodes 3 and 4 ($V_{34}$), separated by a distance $L$:
\begin{equation}
    R_{12;34} = \frac{V_{34}}{I_{12}} = \rho_{L} \left(\frac{L}{wt}\right),
    \label{eq:R_L}
\end{equation}

where $\rho_L$ is the longitudinal resistivity. For measuring the transverse resistance ($R_{T}$), the voltage difference is measured across electrodes 4 and 5 while the current is sent from 1 to 2:
\begin{equation}
    R_{12;45} = \frac{V_{45}}{I_{12}} = \rho_{T}\left(\frac{1}{t} \right)
    \label{eq:R_T}
\end{equation}

As an important note, the functional forms of Eq.~(\ref{eq:R_L}) and Eq.~(\ref{eq:R_T}) are obtained by integrating the differential form of Ohm's law $\vec{J}_{i} = \sigma_{ij} \vec{E}_{j}$. For this calculation, the current density, $\vec{J}$, is assumed to be uniform along the x-direction ($\vec{J} = J \hat{x}$). To achieve such current flow, electrodes 1 and 2 must span the width of the two ends, the sample thickness must be much smaller than its width ($t \ll W$) and length($t \ll L$), and the voltage electrodes must be small. The complication happens when applying the magnetic field. Here, we limit our attention to the case where the magnetic field is applied perpendicular to the transport plane. 

We first focus on the longitudinal resistivity, $\rho_{L}$. When the magnetic field is not applied, we can regard this resistivity to be equal to both $\rho_{xx}$ or $1/\sigma_{xx}$. However, once the perpendicular magnetic field is applied to the sample, these equalities do not in general hold true anymore (i.e., $\rho_{L} \neq \rho_{xx}$ and $\rho_{L} \neq 1/\sigma_{xx}$). In Figure~(\ref{Fig:FEA_LongitudinalResistance}), we simulate the change in $R_{L}$ using finite element analysis as an instructive example. A typical magnetotransport measurement with a length-to-width ratio ($L/W$) ranging from 1 to 5 shows that the change in magnetic field behavior is neither following $\rho_{xx}$ in Eq.~(\ref{eq:rho_xx}) nor $1/\sigma_{xx}$ in Eq.~(\ref{eq:sigma_xx}). Although this distinction may naively appear unrelated to the analysis of the anomalous Hall effect, it becomes important when converting the measured resistance values into ($\sigma_{yx}$). Since obtaining $\sigma_{yx}$ involves a matrix inversion involving ($\rho_{xx}$), it is critical to find $\rho_{xx}$, not an intermediate value between   $\rho_{xx}$ and $1/\sigma_{xx}$.

A natural following up question is whether the transverse resistance is truly an off-diagonal resistivity measurement, as described in Eq.~(\ref{eq:R_T}). This turns out to be more robust than the longitudinal measurement in that there is no ambiguity such as measuring between $\rho_{yx}$ and $\sigma_{yx}$. In general, a transverse measurement is proportional to the off-diagonal resistivity $\rho_{yx}$. However, the naive relation , scaling with the thickness ($R_{T} = \rho_{yx} /t$) is not guaranteed\cite{moelter1998electric} because the boundary conditions mentioned above are not strictly satisfied for non-ideal geometries. In general, we can express the transverse resistance as,

\begin{equation}
    R_{T} = \rho_{yx}\frac{g(L/W)}{t}.
\end{equation}

As shown in our numerical simulations shown in Figure~(\ref{Fig:FEA_LongitudinalResistance})~(b), where the transverse voltage is evaluated at the midpoint of the two sample edges, the measured transverse resistance decreases as the aspect ratio $L/W$ is reduced, corresponding to a geometrical correction factor $g(L/W)<1$. More generally, this correction factor also depends on the positions of the transverse voltage electrodes relative to the current leads. Therefore, a more realistic magnetotransport geometry, such as that shown in the inset of Figure ~(\ref{Fig:FEA_LongitudinalResistance}) (b), will generally be characterized by a different value of $g$.

So far, we have focused on the ordinary Hall effect and several subtleties that arise in magnetotransport measurements. We now consider the case in which an anomalous Hall contribution is also present. In particular, we examine how the ordinary and anomalous Hall conductivities combine and how their coexistence influences the measured longitudinal and transverse responses.

\subsection{Combined Ordinary and Anomalous Hall Transport} 

For most studies of the anomalous Hall effect, the quantity of interest is the anomalous Hall conductivity, $\sigma_{yx}^{AHE}$. Unlike the ordinary Hall contribution, $\sigma_{yx}^{AHE}$ does not have a universal magnetic field dependence. In conventional ferromagnets, for example, it often follows the magnetization,  $\sigma_{yx}^{AHE} \propto M(H)$, and can therefore exhibit features such as saturation and hysteresis. Much recent interest has focused on intrinsic anomalous Hall effects arising from the Berry curvature, $\Omega(k)$, of the electronic bands. Berry phases can also arise from real-space noncoplanar spin textures, giving rise to an additional Hall response commonly referred to as the topological Hall effect. In this work, we do not distinguish between the microscopic origins of these excess Hall contributions, but collectively denote them by $\sigma_{yx}^{AHE}$. 

When an anomalous Hall effect is present, it contributes together with the ordinary Hall effect. Similar to the two channel analysis introduced in SubSection~(\ref{SubSec:Drude}), the two contributions are summed by each matrix element of the conductivity matrix. Therefore, the Hall conductivity have the following form, 

\begin{equation}
\sigma_{yx}^{tot}=\sigma_{yx}^{O}+\sigma_{yx}^{A},
\label{eq:totalHallconductivtiy}
\end{equation} 

where $\sigma_{yx}^{O}$ is the ordinary Hall response. As reviewed in the previous subsection (SubSec.~\ref{SubSec:ResistivityConductivity}), the transverse resistance measurement is proportional to $\rho_{yx}$. From the inversion of conductivity matrix, the measured Hall resistivity becomes

\begin{equation}
    \rho_{yx}^{M} = -\frac{\sigma_{yx}^{O}+\sigma_{yx}^{AHE}}{\sigma_{xx}^2 + (\sigma_{yx}^{O}+\sigma_{yx}^{AHE})^2}.
    \label{eq:TotalHallresistivity}
\end{equation}

In the limit of $|\sigma_{yx}^{O} + \sigma_{yx}^{AHE}|\ll \sigma_{xx}$,
and $\sigma_{yx}^{AHE} \propto M$ in ferromagnetic systems, this converges
to the total Hall resistivity empirically written for ferromagnetic
systems\cite{pugh1930halleffect,pugh1932hallemf},

\begin{equation}
    \rho_{yx}^{M} = R_{O}H + R_{s}\mu_{0}M,
    \label{eq:hall_oa}
\end{equation}

where $R_{O}$ ($\approx -\sigma_{yx}^{O}/\sigma_{xx}^2$) and $R_{s}$ ($\approx -\sigma_{yx}^{AHE}/\mu_{o} M\sigma_{xx}^2$) are the ordinary and anomalous Hall coefficients,
respectively. It is important to note that this limit may not always hold true, and therefore it is safer to consider Eq.~(\ref{eq:TotalHallresistivity}).

Assuming the ordinary magnetotransport can be expressed as a single channel Drude model , and the limiting case of $\sigma_{yx}^{AHE} \ll 1/\rho_{xx}^{O}$ and $\mu B \ll 1$, we find

\begin{equation}
\begin{aligned}
\rho_{xx}^{M}
&= \frac{\sigma_{xx}^{O}}
{(\sigma_{xx}^{O})^2 + (\sigma_{yx}^{O}+\sigma_{yx}^{AHE})^2} \\
&\approx \rho_{xx}^{O}
- (\rho_{xx}^{0})^3
\left[
(\sigma_{yx}^{AHE})^2
+ 2(\sigma_{yx}^{AHE})(\sigma_{yx}^{O})
\right].
\end{aligned}
\label{eq:rho_xx_mixed_AHC}
\end{equation}

where $\rho_{xx}^{O} = 1/ne\mu$ and $\sigma_{yx}^{O} =- sne\mu^2 B/(1+\mu^2 B^2)$. 

The surprising fact is that an ideal $\rho_{xx}$ measurement, which was found to be a constant in the absence of the anomalous Hall conductivity channel becomes field dependent due to the mixing of both $\sigma_{yx}^{AHE}$ and $\sigma_{yx}^{O}$. Also, these contributions appear as even functions of the magnetic field, and therefore a symmetrization process of the data will not remove this contribution. 

In the following sections, we apply the transport framework developed in this section to the analysis of the anomalous Hall conductivity, $\sigma_{yx}^{AHE}$. We first examine SrRuO$_3$ as a benchmark system in which the anomalous Hall response can be identified relatively clearly, and then turn to GdAlSi, where separating the ordinary and anomalous contributions is considerably more subtle and challenging.

\section{S$\textrm{r}$R$\textrm{u}$O$_3$ as a Benchmark for Anomalous Hall Analysis}

\begin{figure}[!htbp]
    \centering
    \includegraphics[width=0.95\columnwidth]{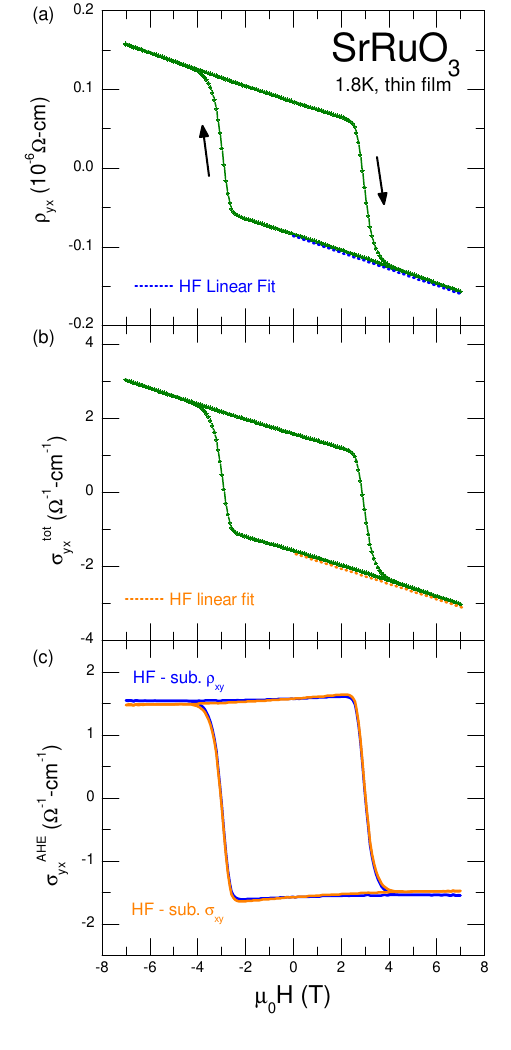}
    \caption{Anomalous Hall conductivity extraction in a SrRuO$_3$ thin film at 1.8 K. (a) Antisymmetrized transverse resistivity, $\rho_{yx}$, with a high-field linear fit used to estimate the ordinary Hall background (shown in blue dotted lines labeled HF linear fit). (b) Total transverse conductivity, $\sigma_{yx}$, with a high-field linear fit (shown in orange dotted lines labeled HF linear fit) used to estimate the ordinary Hall contribution directly in conductivity space. (c) Extracted anomalous Hall conductivity, $\sigma_{yx}^{\mathrm{AHE}}$, obtained from the resistivity-based procedure (blue, labeled HF-sub. $\rho_{yx}$) and the conductivity-based procedure (orange, labeled HF-sub. $\sigma_{yx}$).
}
    \label{Fig:SR113_HighFieldAnalysis}
\end{figure}

In this section, we study the anomalous Hall conductivity extraction process with a benchmark material SrRuO$_3$. This material is an itinerant ferromagnetic metal where it exhibits an anomalous Hall effect (AHE), being one of the most important aspects of this material. The anomalous Hall response is reported to contain a substantial intrinsic Berry curvature contribution originating from the spin-orbit-coupled electronic structure. The magnitude and even the sign can depend sensitively on temperature, scattering, and epitaxial strain\cite{haham2011scaling_SrRuO3, tian2009proper}.

The conventional high-field subtraction can be understood from the phenomenological decomposition introduced in Eq.~(\ref{eq:hall_oa})\cite{haham2011scaling_SrRuO3}. Once the magnetization reaches saturation, $M_{s}$, the anomalous Hall contribution is expected to become approximately field independent, provided that the electronic and scattering properties governing the AHE do not vary appreciably over the fitting range. The high-field Hall resistivity can then be written as $\rho_{yx} \approx R_{0} H + \rho_{yx}^{AHE, Sat.}$, such that the slope gives the ordinary Hall coefficient $R_{O}$, while the zero-field intercept of the high-field fit gives the saturated anomalous Hall resistivity. This assumption is well suited to SrRuO$_3$, where the magnetization and anomalous Hall response have both been observed to reach a saturated high-field regime\cite{tian2009proper}. 

Figure~(\ref{Fig:SR113_HighFieldAnalysis}) shows our Hall measurement and analysis of a SrRuO$_{3}$ thin film. We first antisymmetrize the transverse resistance where the detailed process is detailed in SI III. The antisymmetrized transverse resistivity ($\rho_{T} = \rho_{yx}$), as shown in Figure~(\ref{Fig:SR113_HighFieldAnalysis})~(a), exhibits a clear hysteretic response within a $\pm 4$ T window, while exhibiting a non-hysteretic linear field dependence at higher magnetic field ranges. We fit this high-field linear region to estimate the ordinary Hall contribution and subtract it from the measured $\rho_{yx}^{M}$. This is then converted to the anomalous Hall conductivity with a matrix inversion, as shown in the blue curve (labeled HF-sub. $\rho_{yx}$) in Figure~(\ref{Fig:SR113_HighFieldAnalysis})~(c). An alternative approach is converting the measured longitudinal and transverse resistivity to $\sigma_{yx}^{tot}$ first and fitting at high fields to estimate the ordinary Hall conductivty, $\sigma_{yx}^{O}$, as shown in Figure~(\ref{Fig:SR113_HighFieldAnalysis})~(b). The resulting anomalous Hall conductivity from this method is shown in the orange curve (labeled HF-sub. $\sigma_{yx}$ in Figure~(\ref{Fig:SR113_HighFieldAnalysis})~(c)). The two procedures yield nearly identical results for SrRuO$_3$. 

Our SrRuO$_3$ measurements and analysis demonstrate that the anomalous Hall conductivity can be robustly extracted using two distinct background-subtraction procedures, which yield nearly identical results. In the following section, we show that, in contrast, the same two approaches produce dramatically different estimates of $\sigma_{yx}^{\mathrm{AHE}}$ in GdAlSi. We then discuss the possible origins of this discrepancy and strategies for extracting the anomalous Hall contribution in cases where the ordinary Hall background is not uniquely determined.

\section{Anomalous Hall Analysis in G$\textrm{d}$A$\textrm{l}$S$\textrm{i}$}

As discussed in the Introduction, previous studies have reported both magnetic order\cite{Laha_GdAlSi, gong2024magnetic, Nag_GdAlSi, li2026field, nakano2026perfectly} and a topological electronic band structure\cite{Laha_GdAlSi, Nag_GdAlSi} in GdAlSi, although the details do not have full agreement. When focusing on magnetotransport, there is a weak change in the slope. This change in slope becomes significantly weaker above the Neel temperature at 32 K (see SI II.)\cite{Laha_GdAlSi, Nag_GdAlSi}. In fact, an anomalous Hall effect has also been reported, analyzing this change in slope\cite{Laha_GdAlSi}. In this section, we critically examine the possibility of the anomalous Hall conductivity using several different approaches. 

In our experiment, we use GdAlSi single crystals. Details of the GdAlSi single-crystal growth are provided in the Supplemental Material (see SI I). Both the measured transverse resistance and longitudinal resistance were converted to $\rho_{yx}$ and $\rho_{L}$ by simulating the real transport geometry of the sample using Finite Element Analysis (See SI IV.). We mainly study the 2 K magnetic field sweep data of $\rho_{yx}$ shown in Figure~(\ref{Fig:GdAlSi_HighFieldAnalysis})~(a) using different analysis methods.

\subsection{High-Field Analysis}

\begin{figure}[!htbp]
    \centering
    \includegraphics[width=0.9\columnwidth]{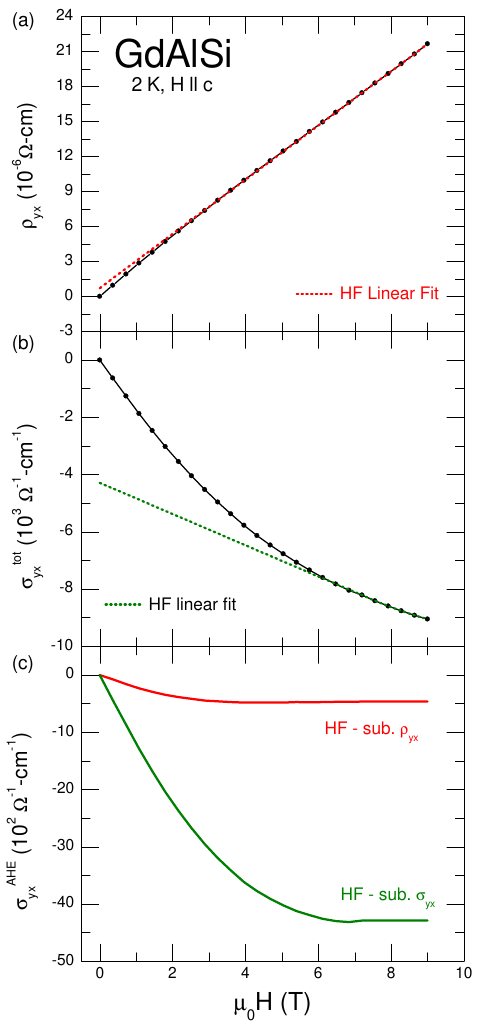}
    \caption{Extraction of the anomalous Hall effect using high-field linear extrapolation. (a) Measured Hall resistivity, ($\rho_{yx}^{M}$), at 2 K for ($H\parallel c$). The red dashed line (labeled HF linear fit) denotes the linear fit to the high-field region extrapolated to lower fields. (b) The total Hall conductivity, ($\sigma_{yx}^{tot}$), calculated from the measured resistances. The green dashed line (labeled HF linear fit) denotes the corresponding high-field linear extrapolation. (c) Estimated anomalous Hall conductivity obtained using two subtraction procedures: 1) HF-sub. $\rho_{yx}$ (in red): subtraction of the high-field linear background from $\rho_{yx}$ prior to conversion to conductivity. 2) HF-sub. $\sigma_{yx}$ (in green): subtraction of the high-field linear background from $\sigma_{yx}$.}
    \label{Fig:GdAlSi_HighFieldAnalysis}
\end{figure}

 The Hall data at 2 K, after converting from measured resistance to resistivity, is shown in Fig.~(\ref{Fig:GdAlSi_HighFieldAnalysis})~(a). We first consider the ordinary Hall effect estimation from the high field data. In contrast to SrRuO$_3$, the change in slope is much more subtle. However, a slight difference in the slope between high fields and low fields exist, emphasized by the high field extrapolation shown in the red dotted lines. This difference in slope becomes much smaller above $T_{N}$, suggesting a possible excess channel origin from the magnetic order. Using the slope from the linear fit at high magnetic field ranges (7.2 - 9 T), then extrapolating to the entire field range (shown in the red dotted line), we estimate the ordinary Hall resistivity. As a comparison study, we also convert the measured resistance values to $\sigma_{yx}^{tot}$, assuming $\rho_{T} = \rho_{yx}$ and $\rho_{L} = \rho_{xx}$. Unlike $\rho_{yx}^{M}$ in Figure~(\ref{Fig:GdAlSi_HighFieldAnalysis})~(a), $\sigma_{yx}^{tot}$ in Figure~(\ref{Fig:GdAlSi_HighFieldAnalysis})~(b), shows a clear slope change. We also use the slope from the linear fit at high magnetic field ranges (7.2 T - 9 T) and extrapolate the resulting fit over the entire field range.

Unexpectedly, we find that the resistivity-based and conductivity-based ordinary Hall effect subtraction procedures yield dramatically different results. The ordinary Hall resistivity estimated from the high-field linear fit in Fig.~(\ref{Fig:GdAlSi_HighFieldAnalysis})~(a) was first converted to an ordinary Hall conductivity, and then subtracted from the total Hall conductivity, calculated by $\sigma_{yx}^{tot} = - \rho_{yx}^{M}/(\rho_{xx}^2 + (\rho_{yx}^{M})^2)$. The anomalous Hall conductivity obtained using this procedure is shown by the red curve in Fig.~(\ref{Fig:GdAlSi_HighFieldAnalysis})~(c). Alternatively, one can fit the high-field region directly in the Hall-conductivity representation, as shown in Fig.~(\ref{Fig:GdAlSi_HighFieldAnalysis})~(b), and subtract this fitted ordinary Hall conductivity from the total Hall conductivity. The resulting anomalous Hall conductivity is shown by the green curve in Fig.~(\ref{Fig:GdAlSi_HighFieldAnalysis})~(c).

One might naively expect that the anomalous Hall conductivity extracted from the conductivity-based fit is more reliable than the resistivity-based fit, since Eq.~(\ref{eq:totalHallconductivtiy}) provides the more fundamental description of how the ordinary and anomalous Hall contributions combine, whereas Eq.~(\ref{eq:hall_oa}) is derived for a limited case. However, we find that this is not the case and we detail the reason in the following.

\begin{figure}[!htbp]
    \centering
\includegraphics[width=0.98\columnwidth]{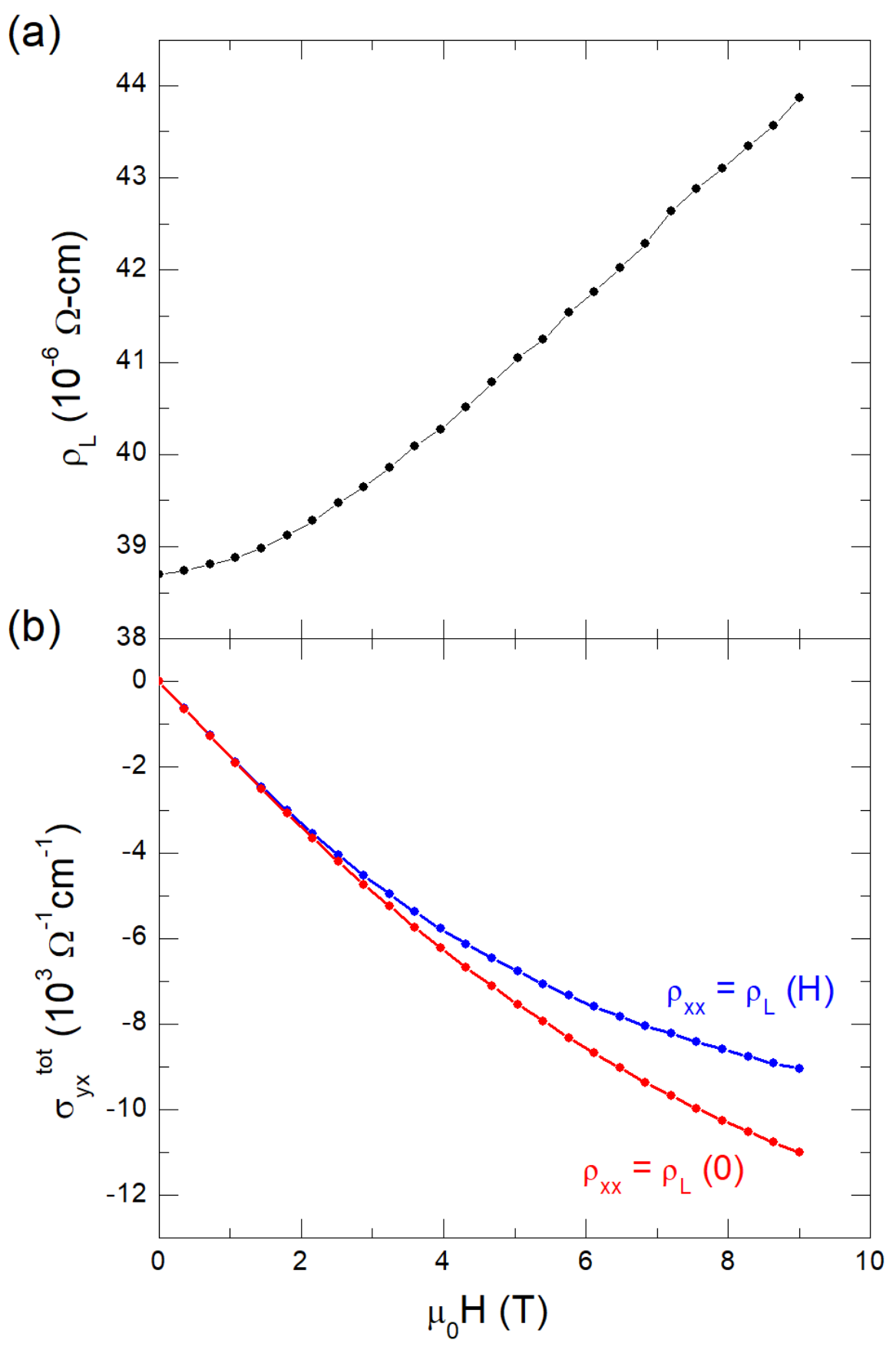}
     \caption{Effect of the field-dependent longitudinal resistivity on the extracted Hall conductivity. (a) Measured longitudinal resistivity, $\rho_{L}^{M}$, of GdAlSi at 2 K for $H\parallel c$. (b) Comparison of the total Hall conductivity obtained by direct inversion by assuming the measured longitudinal resistivity $\rho_{L}$ is $\rho_{xx}$ (in blue) and that assuming only the zero field $\rho_{L}$ is maintained along the entire range (in red). This uncertainty in $\rho_{xx}$ contributes to the discrepancy in the extracted anomalous Hall conductivity.}
    \label{fig:reason_discrepency}
\end{figure}

As shown in Figure~(\ref{fig:reason_discrepency})~(a), the measured longitudinal resistivity, $\rho_{L}$, increases with increasing magnetic field. In converting to the total Hall conductivity, we assumed that this field dependence directly represents $\rho_{xx} (H)$. However, as shown in Fig.~(\ref{Fig:FEA_LongitudinalResistance})~(a), this is not always the case. When the transport geometry measures $\rho_{L} = 1/\sigma_{xx}$ the field dependence is soley due to the orbital effect (Lorentz force). On the other hand, if $\rho_{L} = \rho_{xx}$, this is a constant when the anomalous Hall conductivity is absent, but also becomes field dependent, as described in Eq.~(\ref{eq:rho_xx_mixed_AHC}), in the presence of the anomalous Hall conductivity.

Figure~(\ref{fig:reason_discrepency})~(b) illustrates the consequence of total conductivity matrix conversions of the two limiting cases of $\rho_{L}$. The blue curve assumes $\rho_{L}(H) = \rho_{xx}$, whereas the red curve assumes a field-independent longitudinal resistivity $\rho_{L}(0)=\rho_{xx}$. Although both are constructed from the same transverse resistance data, the resulting total Hall conductivities exhibit substantially different high-field slopes. Consequently, the ordinary Hall contribution inferred from a conductivity-based analysis depends strongly on how the measured longitudinal response is interpreted. 

In order to avoid the uncertain interpretation of $\rho_{L}$, we propose an alternative approach of finding $\sigma_{yx}^{AHE}$. We first find the Hall mobility from the high field resistivity slope and zero field longitudinal resistivity (interpreted as $\rho_{L} (0) = 1/ne\mu$), 

\begin{equation}
    \mu = \frac{1}{\rho_{L}(0)}\left| \frac{d\rho_{yx}}{dB}\right|
    \label{eq:Hall_mobility}
\end{equation}

We then use this mobility to estimate the orbital field dependence of the longitudinal conductivity,

\begin{equation}
    \sigma_{xx}^{O}(B) = \frac{ne\mu}{1+\mu^2B^2} = \frac{1}{\rho_{L}(0)}\frac{1}{1+\mu^2B^2}
\end{equation}

With $\sigma_{xx}^{O}(B)$ determined in this manner, the anomalous Hall conductivity can be obtained from the measured Hall resistivity in Eq.~(\ref{eq:TotalHallresistivity}). This can be expressed as

\begin{equation}
\begin{aligned}
0 = \rho_{yx}^{M}\Big[
    &(\sigma_{yx}^{AHE})^2
    + 2\sigma_{yx}^{O}\sigma_{yx}^{AHE} \\
    &+ (\sigma_{xx}^{O})^2
    + (\sigma_{yx}^{O})^2
    \Big]
    + \sigma_{yx}^{AHE}
    + \sigma_{yx}^{O}.
\end{aligned}
\label{eq:total_conductivity_solving}
\end{equation}

Equation~(\ref{eq:total_conductivity_solving}) is quadratic in the anomalous Hall conductivity and therefore yields two mathematical solutions. One solution corresponds the case in the $|\sigma_{yx}^{tot}| < \sigma_{xx}$ limit, while the other corresponds to the $|\sigma_{yx}^{tot}| > \sigma_{xx}$ case. Since the measured transverse response is much smaller than the longitudinal response in the present experiment, we select the $|\sigma_{yx}^{tot}| < \sigma_{xx}$ solution. 

\begin{figure}[!htbp]
    \centering
\includegraphics[width=1.0\columnwidth]{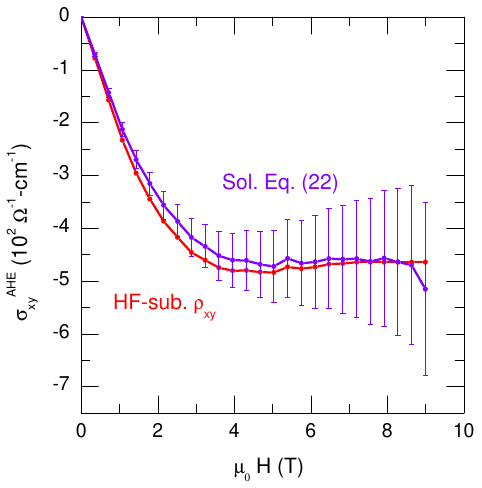}
     \caption{Comparison of anomalous Hall conductivity extracted using two high-field analysis methods. The red curve (HF-sub. ($\rho_{yx}$)) is obtained by subtracting a high-field linear background from ($\rho_{yx}$) prior to conversion to conductivity. The violet curve is obtained by solving for $\sigma_{yx}^{AHE}$ from Eq.~(\ref{eq:total_conductivity_solving}), after $\sigma_{xx}^{O}$ has been modeled from the Hall mobility. The error bars show the variation in the extracted ($\sigma_{yx}^{AHE}$) when the Hall mobility is varied by $1.5\%$.}
    \label{fig:DrudeModeling}
\end{figure}

The $\sigma_{yx}^{AHE}$ found from this Drude modeling of $\sigma_{xx}^{O}(B)$ is shown in the violet curve of Figure~(\ref{fig:DrudeModeling}). Notably, it agrees well with the anomalous Hall conductivity obtained by first subtracting the high-field linear background from $\rho_{yx}^{M}$ and subsequently converting the remaining Hall response to conductivity, rather than by directly fitting the converted $\sigma_{yx}^{\mathrm{tot}}$. We note that an accurate estimation of mobility from Eq.~(\ref{eq:Hall_mobility}) is a critical step for this analysis process. The error bars in Figure~(\ref{fig:DrudeModeling}) indicate how much the estimated $\sigma_{yx}^{AHE}$ changes when the mobility value is off by 1.5$\%$. We find the results are more vulnerable at higher magnetic fields (i.e., when $\mu B$ is larger). Lastly, we note that the overall better agreement with the HF-sub. $\rho_{yx}$ than HF-sub. $\sigma_{yx}$ does not suggest that the $\rho_{yx}$-based fit procedure is always more reliable in general. At low fields (0 to 2 T) in Fig.~(\ref{fig:DrudeModeling}), the two results show a small but noticeable discrepancy. Again, this is caused by the incorrect estimation of $\rho_{xx}$ in the red curve (using HF-sub $\rho_{yx}$ method). We show in the following section that this effect becomes greatly amplified when estimating the ordinary Hall conductivity using the low field data. 

From this exercise, we learn that both the total Hall and the ordinary Hall conductivity can be incorrectly estimated because the measured longitudinal response, $\rho_L$, is improperly identified as either $\rho_{xx}$ or $1/\sigma_{xx}$ and was used for the resistivity to conductivity matrix conversion. In particular, the field dependence of the converted total Hall conductivity, $\sigma_{yx}^{\mathrm{tot}}$, will have an incorrect curvature.  This uncertainty exists both in the resistivity-based and conductivity based analysis. Even worse, if the ordinary Hall conductivity, $\sigma_{yx}^{\mathrm{O}}$, is estimated from the slope of that wrongly estimated total Hall conductivity (i.e. conductivity-based analysis), it will introduce an extra error for estimating $\sigma_{yx}^{\mathrm{AHE}}$. 

It is instructive to compare this situation with SrRuO$_3$ [Fig.~(\ref{Fig:SR113_HighFieldAnalysis})]. For SrRuO$_3$, $\mu B \ll 1$, such that $\rho_{xx}^{0} \approx 1/\sigma_{xx} \approx 1/ne\mu$, and the distinction between these quantities is negligible. In GdAlSi, however, we estimate $\mu B \sim 0.7$ at 10~T. In this regime, $1/\sigma_{xx}$ can acquire a substantial orbital field dependence even when the intrinsic Drude resistivity, $\rho_{xx}^{0}=1/ne\mu$, remains field independent. Therefore, attributing the measured field dependence of $\rho_L$ entirely to $\rho_{xx}$ can introduce a significant error in the extracted Hall conductivity.

Taken together, the high-field analysis shows that the extracted anomalous Hall conductivity is highly sensitive to the interpretation of the measured longitudinal response. When $\mu B$ is not negligible, correctly determining $\sigma_{xx}^{O}$, including its field dependence, becomes highly important for the reliable anomalous Hall conductivity analysis. 

\subsection{Low-Field Analysis}

\begin{figure}[!htbp]
    \centering
    \includegraphics[width=0.98\columnwidth]{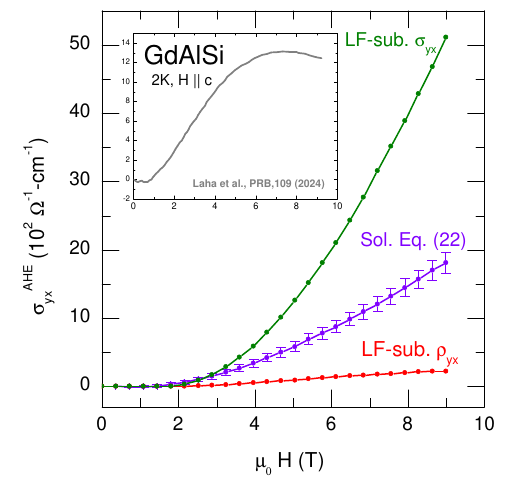}
     \caption{Comparison of anomalous Hall conductivity extracted using three low-field analysis methods. The red curve (LF-sub. $\rho_{yx}$) is obtained by subtracting a low-field linear background from ($\rho_{yx}$) prior to conversion to conductivity. The green curve (LF-sub. $\sigma_{yx}$) is obtained by subtracting a low field linear background from $\sigma_{yx}^{tot}$. The purple curve is obtained by solving for $\sigma_{yx}^{AHE}$ from Eq.~(\ref{eq:total_conductivity_solving}), after $\sigma_{xx}^{O}$ has been modeled from the Hall mobility. The error bars show the variation in the extracted ($\sigma_{yx}^{AHE}$) when the Hall mobility is varied by $\pm 0.75\%$. Inset: Previous $\sigma_{yx}^{AHE}$ report by Laha $et$ $al$.\cite{Laha_GdAlSi}, using a low-field linear background subtraction.}
    \label{fig:LowFieldAHC_Extraction}
    \end{figure}

Although the ordinary Hall background is commonly estimated from the high-magnetic-field regime, there are cases when the low-field regime is used instead\cite{Laha_GdAlSi}. This is when the anomalous Hall channel is absent or sufficiently smaller than the ordinary Hall channel at low magnetic fields. When the magnetic field modifies the magnetic or electronic structure, the associated anomalous Hall conductivity can be zero or sufficiently small at low fields and later turn on at higher fields. For example, MnSi under pressure, the excess Hall contribution only appears within a finite magnetic field range\cite{nagaosa2010anomalous}. 

In GdAlSi, Laha $et$ $al$. adopted the low-field linear Hall response as the background that needs to be subtracted. This choice appears to be motivated by the more linear behavior of $\rho_{yx}$ at low fields\cite{Laha_GdAlSi}. They reported a very large $\sigma_{yx}^{AHC} \approx 1310 \ \Omega^{-1} \textrm{cm}^{-1}$ (also shown in the inset of Figure~(\ref{fig:LowFieldAHC_Extraction})), which they explain that it is more consistent with the Weyl-band Berry curvature origin than the non-trivial real-space spin-texture origin\cite{Laha_GdAlSi}. Motivated by this prior study, in this section, we critically examine the anomalous Hall conductivity using the low-field Hall response as the ordinary Hall background.

We apply the same methodology introduced in the previous section, except that the ordinary Hall contribution is estimated from the low-field regime (ranging from 0 to 1.8 T). In Figure~(\ref{fig:LowFieldAHC_Extraction}), we compare the anomalous Hall conductivity extracted using three approaches: subtracting a linear low-field background from $\rho_{yx}$, subtracting a linear low-field background from $\sigma_{yx}^{tot}$,  and solving for $\sigma_{yx}^{AHE}$ using Eq.~(\ref{eq:total_conductivity_solving}). The three methods yield substantially different results, indicating that the extracted anomalous Hall conductivity is highly sensitive to the analysis procedure. 

The discrepancy largely originates from the interpretation of $\rho_{L}$. When calculating $\sigma_{yx}^{tot}$, a poor estimation of $\sigma_{xx}$ from $\rho_{L}$ leads to conversion errors. These errors become amplified at higher magnetic fields as $\mu B$ increases. While this error was also present in the previously discussed high-field subtraction, the same error existed in both the total and ordinary Hall conductivities. Because both exhibited the same magnitude of transverse resistivity, their curvatures effectively canceled out, leaving a nearly constant value at high fields.

Another key cause of this discrepancy is the wrong assumption in the LF-sub. $\sigma_{yx}$ method. When performing a fit from $\sigma_{yx}^{tot}$ at low fields and extrapolating to the entire field range, we are assuming the ordinary Hall conductivity remains linear at high magnetic fields. However, recall the behavior of the ordinary Hall conductivity
 
 \begin{equation}
     \sigma_{yx}^{O} = -s\frac{ne\mu}{1+\mu^2 B^2} \mu B
 \end{equation}
 
 At low fields ($\mu B \ll 1$), the field dependence in the denominator is negligible, making the linear approximation valid. However, at high magnetic fields, this term becomes significant, and therefore the linear extrapolation is incorrect.

In this section, we have shown that the low-field analysis reveals two sources of uncertainty: 1) the conversion of the measured $\rho_{L}$ to $\sigma_{xx}$, and 2) the assumption that the low-field estimation of $\sigma_{yx}$ can be extrapolated linearly to higher fields. 

\subsection{Single Channel Ordinary Hall Interpretation Without an Anomalous Hall Contribution}

\begin{figure}[!htbp]
    \centering
    \includegraphics[width=0.98\columnwidth]{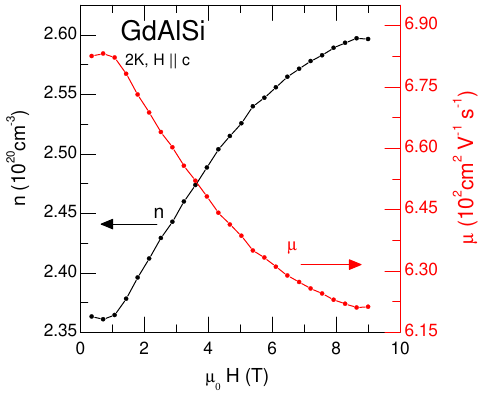}
     \caption{Single-channel ordinary Hall analysis of GdAlSi. Magnetic field dependence of the carrier concentration (n, left axis) and carrier mobility ($\mu$, right axis) for GdAlSi measured at 2K with $H \parallel c$.}
    \label{fig:Figure 7}
\end{figure}

As an alternative interpretation, we consider whether the measured Hall response of GdAlSi can be explained entirely by the ordinary Hall effect, with $\sigma_{yx}^{AHE} = 0$. Within a single-channel Drude picture, this would require the non-linearity of the measured transverse response to arise solely from the field dependence of the ordinary conductivity tensor. We therefore examine what evolution of the carrier density and mobility would be required to reproduce the measured Hall signal and whether such an interpretation is consistent with the simultaneously measured longitudinal magnetotransport.

Since the Hall slope is not perfectly linear, the transport paramters can no longer remain field independent. Over the field range, the inferred carrier density increases about 8 $\%$ from roughly $2.39 \times10^{20}$ cm$^{-3}$ to $2.58 \times10^{20}$ cm$^{-3}$ , while the inferred mobility decreases a comparable amount from about 685 cm$^2$/V-sec to 620 cm$^2$/V-sec. We emphasize that this analysis does not necessarily imply that $n$ and $\mu$ physically evolve in this manner. Rather, it demonstrates that a field-independent single-channel Drude model is insufficient to account for the measured Hall response in the absence of an anomalous Hall contribution. Similar field-dependent transport parameters have also been reported in other f-electron systems. For example, magnetotransport analysis of SmB$_6$ found an increase in carrier density with magnetic field accompanied by a decrease in mobility.\cite{WolgastSmB6}.

\subsection{Two Channel Ordinary Hall Interpretation Without an Anomalous Hall Contribution}

\begin{figure}[!htbp]
    \centering
    \includegraphics[width=0.98\columnwidth]{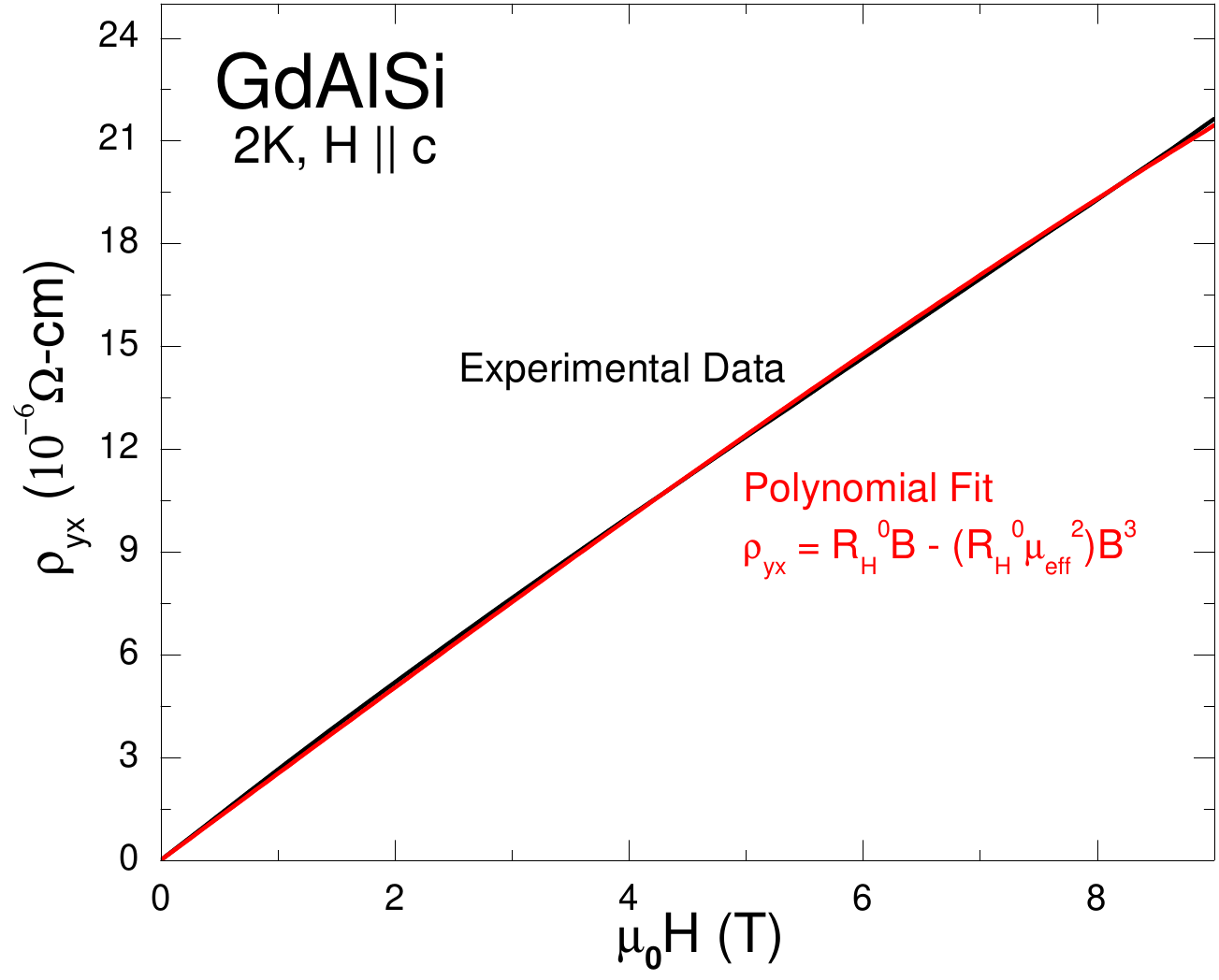}
     \caption{Two-channel ordinary Hall analysis of GdAlSi. The measured transverse resistivity, $\rho_{yx}$ at 2 K shown in black solid lines and the polynomial fit is shown in the red solid line. The fit results in coefficients of $R_{o}^{H} = 2.53 \times 10^{-6} \pm 9.72\times 10^{-9}$, and $-R_{o}^{H}\mu_{eff}^2 = 1.75\times 10^{-9} \pm 1.76\times 10^{-10}$.}
    \label{fig:Figure 8}
\end{figure}

In the previous subsection, we showed that attributing the measured transverse resistance using a single channel ordinary Hall effect, without an anomalous Hall contribution, requires the carrier density and mobility to evolve with magnetic field. Here, we consider an alternative scenario in which two ordinary transport channels contribute to the ordinary Hall effect while the anomalous Hall contribution is still absent. This scenario is particularly relevant since it can in general exhibit a nonlinear behavior even when the carrier density and mobility remain field independent. However, a two channel analysis requires four fitting parameters: two carrier densities and two mobilities. This makes the data fitting analysis difficult to constrain. 

Instead we consider the low-field expansion of the two-channel ordinary Hall response, introduced in Eq.~(\ref{Eq:EffectiveTwoHall})

\begin{equation}
    \rho_{yx} = R_{H}^{0}B[1-(\mu_{eff} B)^2+\mathcal{O}((\mu_{eff} B)^4)].
\end{equation}

Figure~(\ref{fig:Figure 8}) shows the resulting fit to the measured transverse resistance at 2 K. The fit weakly matches with the experimental data, and produces a mobility of 264 cm$^2$/V$\cdot$ s. This value does not quantitatively agree with the Hall mobility estimated in the previous section (685 cm$^2$/V-sec). Including higher-order terms may improve the numerical fit, but would not uniquely constrain the underlying two-channel transport parameters.

Although this analysis does not establish that magnetotransport of GdAlSi can be fully explained by the multichannel Drude model, it demonstrates that attributing the non-linearity of the Hall response to the ordinary Hall effect is also non-trivial. The gradual temperature evolution of the slope and curvature could, within such a picture, arise from temperature-dependent mobilities and/or carrier densities. Consequently, the nonlinear Hall response alone is insufficient to uniquely distinguish a multichannel ordinary Hall contribution from an anomalous Hall contribution.

In summary, the absence of an anomalous Hall conductivity cannot be ruled out based on the magnetotransport data alone. However, confirming this scenario requires a better understanding of the transverse Hall curvature and the observed mobility discrepancy.
\section{Conclusion}

In summary, we have examined the extraction of anomalous Hall conductivity in SrRuO$3$ and GdAlSi. Using SrRuO$3$ as a benchmark, we first show that different extraction procedures give consistent anomalous Hall conductivities when the ordinary Hall background is well defined and the longitudinal transport produces only a small correction to the conductivity conversion. In contrast, for GdAlSi, different interpretations of the ordinary Hall background and longitudinal transport can lead to substantially different estimates of $\sigma_{yx}^{\mathrm{AHE}}$. The low- and high-field analyses further demonstrate that identifying a linear portion of $\rho_{yx}(B)$ with the ordinary Hall response is not, by itself, always sufficient to uniquely determine the anomalous Hall conductivity.

Several practical considerations for anomalous Hall conductivity analysis are identified in our study, which we believe should be taken into account in future studies of excess Hall transport in quantum materials:

\begin{itemize}
    \item The measured longitudinal resistance should not automatically be identified with either $\rho_{xx}$ or $1/\sigma_{xx}$. The disparity between the two becomes increasingly important as $\mu B$ departs from the $\mu B \ll 1$ limit, even before $\mu B$ approaches unity.

    \item Ordinary Hall background subtraction performed in the resistivity and conductivity is not generally equivalent because tensor inversion couples the longitudinal and transverse components. Again, the disparity between the two procedures becomes increasingly important as $\mu B$ departs from the $\mu B \ll 1$ limit, even before $\mu B$ approaches unity.

    \item The robustness of an extracted anomalous Hall conductivity should be tested against physically reasonable models of both the ordinary Hall and longitudinal transport channels.

    \item Nonlinear Hall transport should not necessarily be attributed to an anomalous Hall contribution, since multichannel ordinary Hall transport can also produce nonlinear magnetic-field dependence.

\end{itemize}

For future studies, these considerations will become particularly critical when the extracted anomalous Hall conductivity is used to estimate an intrinsic Berry-curvature contribution. For example, in Weyl semimetals, the intrinsic anomalous Hall conductivity contains information about the momentum-space separation and chirality of the Weyl nodes. An inaccurate extraction of the anomalous Hall conductivity can therefore lead to an incorrect quantitative interpretation of the underlying Berry curvature and electronic structure. More generally, our results emphasize the importance of consistency checks in the extraction of anomalous and topological Hall responses, particularly in materials where ordinary Hall transport, magnetoresistance, and excess Hall contributions evolve over comparable magnetic-field scales.
\section{Acknowledgment}

We acknowledge support from the Texas Tech University startup package. We thank Michael S. Fuhrer, Wade DeGottardi, Myounghwan Kim, and Mahdi Sanati for useful discussions.

\bibliography{bibliography}  

\end{document}